\documentclass[12pt]{elsart}
\usepackage{graphicx}
\usepackage{color}
\usepackage{amsmath}
\usepackage{bm}
\begin{document}
\begin{frontmatter}
\title{Dual Magnetic Separator for TRI$\mu$P}

\author{G.P.A. Berg\thanksref{GPAB}\thanksref{Present address}},
\author{O.C. Dermois},
\author{U. Dammalapati},
\author{P. Dendooven}
\author{M.N. Harakeh},
\author{K. Jungmann},
\author{C.J.G. Onderwater},
\author{A. Rogachevskiy},
\author{M. Sohani},
\author{E. Traykov},
\author{L.Willmann}, and
\author{H.W. Wilschut}

\address{Kernfysisch Versneller Instituut, Zernikelaan 25,
9747 AA Groningen, The Netherlands}

\thanks[GPAB]{Corresponding author,
Fax: +1-574-631-5952, E-mail: gpberg@bergs.com} {\thanks[Present
address]{Present address: Dept. of Physics, University of Notre
Dame, Indiana, USA}}

\begin{abstract}
The TRI$\mu$P facility, under construction at KVI, requires the
production and separation of short-lived and rare isotopes. Direct
reactions, fragmentation and fusion-evaporation reactions in
normal and inverse kinematics are foreseen to produce nuclides of
interest with a variety of heavy-ion beams from the
superconducting cyclotron AGOR. For this purpose, we have designed,
constructed and commissioned a versatile magnetic separator that
allows efficient  injection into an ion catcher, i.e., gas-filled
stopper/cooler or  thermal ionizer, from which a low energy
radioactive beam will be extracted.

The separator performance was tested with the production and clean
separation of $^{21}$Na ions, where a beam purity of 99.5\%
could be achieved. For fusion-evaporation products, some of the
features of its operation as a gas-filled recoil separator were
tested.

{\it PACS: 07.55.-w; 
           07.55.+h; 
           29.30.-h; 
           41.85.-p; 
           41.75.-i; 
           25.70.Mn; 
           25.70.-z; 
        \\}

{\it Keywords: Magnetic separator, Gas-filled separator, Secondary
radioactive isotopes \\}

\end{abstract}

\end{frontmatter}

\section{Introduction}

Rare and short-lived radio isotopes are of interest because of
their nuclear properties. They offer unique possibilities for
investigating fundamental physical symmetries, for applied
physics, and for nuclear structure studies \cite{HA04}. The main
motivation for investigating fundamental symmetries is to improve
limits for the validity of the Standard Model, which can be
inferred from high-precision measurements. Such low energy
experiments are complementary to searches for new physics in
High-Energy physics experiments. In particular, high accuracy can
be achieved, when suitable radioactive isotopes can be cooled and
stored in atom or ion traps \cite{TU00,JU02,WI03,JU05}.

With this aim the TRI$\mu$P (Trapped Radioactive Isotopes:
$\mu$icrolaboratories for fundamental Physics) facility at the
Kernfysisch Versneller Instituut(KVI) in Groningen, The Netherlands,
was proposed and funded in order to provide a state of the art user
facility for such high-precision studies \cite {BE03,BE03a}. While
the magnetic separator of TRI$\mu$P is the main topic of this article,
we will briefly describe the complete facility consisting of several
major subsystems in order to put the use of the separator into
perspective.

A heavy-ion beam with a maximum magnetic rigidity of 3.6 Tm from
the superconducting cyclotron AGOR \cite{SC99} is used to produce
a variety of short-lived isotopes using very different reaction
mechanisms, such as fragmentation, charge-exchange reactions and
fusion-evaporation. Primary beams ranging from protons to lead are
available. In order to direct most of the reaction products into a
relatively narrow forward-angle cone, the technique of inverse
kinematics is used. This applies e.g. to charge-exchange reactions
on a gaseous hydrogen target and fusion-evaporation on
light-element solid targets. The maximum available beam energy is
about 95 MeV/nucleon for fully stripped, lighter heavy ions
($N=Z$). The energy of heavy ions is restricted by the maximum
charge state that can be reached with the ion source of the
cyclotron. The heaviest ion accelerated with AGOR was $^{208}$Pb
with an energy of 8.4 MeV/nucleon. In the various reactions, the
beam and the reaction products travel together in a narrow cone in
forward direction with the emittance of the products typically
much larger than the beam emittance. The products and beam need to
be separated efficiently, as will be described later.

At the exit of the separator, the produced isotopes have a
considerable momentum corresponding to a maximum magnetic rigidity
of about 3.0 Tm. Here, particles can already be stopped and
measured in e.g. Si detectors to measure certain properties. For
the program on fundamental physics a low-energy secondary beam
must be produced. For this the particles will first be slowed in a
degrader and stopped in a gas-filled ion catcher or a thermal
ionizer. They will be extracted as singly charged ions and guided
into a low pressure He-filled RFQ-Cooler buncher system and
trapped in the last stage of the RFQ, which functions as a Paul
trap. This will allow to produce a low energy bunched beam with a
sufficiently small emittance and an energy of a few keV. This beam
can be guided to one of several experimental setups. One of these
is a Magneto-Optical Trap (MOT) assembly, where the actual
measurements will take place.

The primary goal governing the design concept and the ion optics
was to achieve an optimal separation of the wanted isotopes from
other reaction products and the primary beam. Clearly two
different types of reactions are foreseen. One involving the
production of fast light isotopes from fragmentation or
charge-exchange reactions, the other the production of slow heavy
isotopes in fusion-evaporation reactions. The magnetic separator
was designed to allow two modes, which we will refer to as the
``{\it Fragmentation Mode}'' and the ``{\it Gas-filled Mode}'' for
the two types of reactions, respectively. In the first mode, fully
stripped reaction products can be separated; in the second mode
the charge-state distribution of partly stripped reaction products
can be collapsed dynamically onto a single effective charge state
with a suitable gas filling.

\section{Design of the Dual Magnetic Separator}

At the start of the design of the magnetic separator a list of
criteria was established to develop a concept for the separator
and to specify the design parameters.
\begin{enumerate}
\item Production of a range of short-lived light (e.g. $^{21}$Na)
to heavy radio isotopes (e.g. $^{213}$Ra), using heavy-ion beams
from the AGOR cyclotron. The maximum energy is given by the
operating diagram of the cyclotron and the maximum magnetic
rigidity of 3.6 Tm of the beam line system. Beam stops should be
designed for a maximum of $\approx$ 1 kW beam power dissipation.

\item In the {\it Fragmentation Mode} an efficient collection of
all desired reaction products in an achromatic focal plane, within
a beam spot of the order of 2~cm total transverse dimensions.

\item Effective separation of reaction products and beam, with a
desired suppression factor $\geq 10^{7}$.

\item Rejection of undesired reaction products for a clean
secondary beam, for experiments in the focal plane. (i.e. without
the additional clean-up in the low energy section or in traps.)

\item Optimal and cost efficient use of the existing infrastructure
of the KVI experimental facilities.
\end{enumerate}
\begin{center}
\begin{figure}
\begin{center}
\includegraphics[width=\textwidth,angle=0]{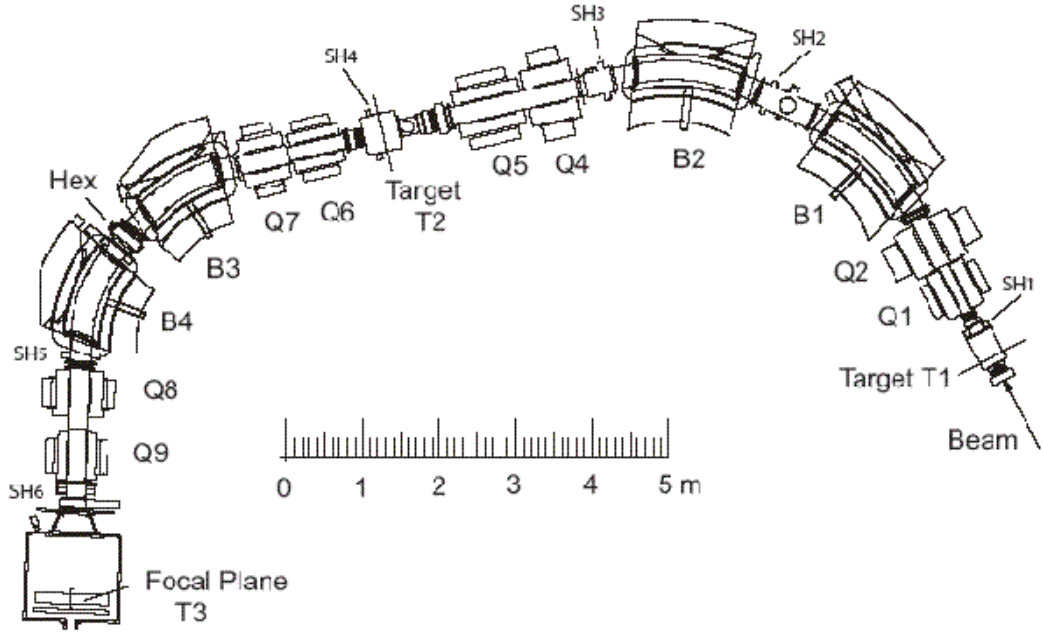}
\caption{Layout of the TRI$\mu$P dual function separator as it is located in the floorplan of the building. North is up. The notations refer to magnetic bending dipoles B, quadrupoles Q, slit systems SH and a hexapole HEX. For further details see text.}
\label{fig:Layout}
\end{center}
\end{figure}
\end{center}

Based on these considerations the combined magnetic separator with
the layout shown in Fig. \ref{fig:Layout} was designed. It is based
on concepts used in previous magnetic fragment \cite
{AN87,SH91,BA02,TA02,KU92,DA92,CO92,RO99,AU02,CO03} and gas-filled
\cite{PA89} separators. It is a special feature of the TRI$\mu$P
separator to combine both concepts in one single device so that it
can operate in either the {\it Fragmentation Mode} or the
{\it Gas-filled Mode}. The main design parameters are summarized in
Table \ref{design parameters}.

\begin{table}
\caption{Design parameters of the TRI$\mu$P magnetic separator.}
\label{design parameters}\vspace {\baselineskip}
\begin{center}
\begin{tabular}{|l|c|c|}
\hline
& Fragment Separator & Gas-filled Separator \\
\hline
Beam rigidity   B$\rho$ &3.6 Tm (Beam line)     &3.6 Tm
({\it Section 1})     \\
Product rigidity B$\rho$        &3.0 Tm ({\it Section 1} and
{\it 2})       &3.0 Tm ({\it Section 2})\\
Solid angle, vert., horiz.      &$\pm$30 mrad   &$\pm$ 30 mrad\\
Momentum acceptance     &$\pm$ 2.0 \%&$\pm$     2.5     \%\\
Resolving Power p/dp &$\approx$ 1000& $\approx$ 2000
(no gas filling)\\
Momentum dispersion &3.9 cm/$\%$&8.0 cm/$\%$\\
Bending radius &220 cm&180 cm\\
\hline
\end{tabular}
\end{center}
\end{table}

The complete magnet system consists of  a total of four dipole
(B1 -- B4) and eight quadrupole magnets (Q1 -- Q9). Q3 originally
foreseen between B1 and B2 in the first designs could be omitted
as discussed in section 3. There are three target chambers, the
one referred to as ``Target T1'' is the target chamber used in
{\it Fragmentation Mode}. The chamber ``Target T2'' is located
at the intermediate focal plane in that mode. It serves as the
target chamber in the {\it Gas-filled Mode}. The third chamber
``Focal Plane T3'' is positioned at the end of the separator.
It is used for measurements in the focal plane. It can be replaced
by the ion catcher. We will refer to the section between T1 and T2
as {\it Section 1} and the section from T2 to T3 as {\it Section 2}.
The provision of dipole doublets in each section instead of one
larger single dipole has several advantages. The additional dipole
edges allow for increased ion-optical control of quadrupole and
higher order corrections. Access spaces are needed in the first
section, between B1 and B2, for slits, scrapers and diagnostic
systems and in {\it Section 2} for a variable hexapole (HEX) for
2$^{nd}$ order corrections.

\subsection{Fragment-Separator Mode}

In {\it Fragmentation Mode}, {\it Section 1} of the separator
focuses the desired isotopes that are created in the object point
(T1) in the intermediate focal plane (T2) with a momentum dispersion
of 3.9 cm/$\%$. The dispersion created by dipoles B1 and B2 allows
the separation of isotopes and beam with different rigidities. For
this purpose slit systems SH2, 3, 4  were installed at the locations
indicated, allowing to stop the beam at various positions depending
on its momentum difference relative to the reaction products that
enter {\it Section 2}. Access via additional ports to install
shields and linings when the beam cannot be intercepted is also made
possible. All spacers that define the gaps of the dipoles are
protected from the direct beam by insulated spacers that allow
current readout for monitoring purposes.

In {\it Section 2} the dispersion is reversed. This provides a
nearly achromatic image in the final focal plane so
that all accepted ions of an isotopes are focused
in the entrance aperture of an ion catcher at that location and
the angular dispersion at T3 is nearly zero. Although this
magnetic analysis only allows the separation of particles with
different rigidities, this may be sufficiently selective when
further purification of the beam follows in e.g. an optical trap.
When the contribution of undesired particles is too high, or in
the case of decay studies with stopped particles in the focal
plane, additional purification is necessary.

To separate different nuclides with the same rigidity a simple and
effective method \cite{AN87,SH91,DU86,SC87} is available that
works well in our energy region. The energy loss of projectiles
passing through a  degrader depends on their atomic numbers Z and
velocities v ($\Delta$E $\propto$ Z$^2$/v$^2$). This can be used
to remove most of the ambiguities of the magnetic separation as
the rigidities will differ after passing the degrader. We used a
flat degrader at T2 to achieve further purification of the desired
isotope  and a slit system in the final focal plane. A degrader
with constant thickness along the intermediate focal plane in T2,
however, will perturb the achromaticity of the system. Therefore,
when the accepted momentum range is large, a degrader in the form
of a ``wedge'' can be used to maintain achromaticity. This may be
done by bending the foil along an appropriate curve.

\subsection {Gas-filled Separator Mode}

The separation concept  of the {\it Fragmentation Mode} described
above does not work for heavy particles at low energies. Here ions
emerge from the production target with a wide charge-state
distribution. The magnetic rigidity B$\rho$, being proportional
to the momentum $p$ and  inversely proportional to the atomic
charge $q$ is not a good measure of the momentum anymore. In
particular, the charge distribution of the beam can cause serious
problems. Even if a charge-state fraction is relatively small, the
number of beam particles can still be orders of magnitude larger
than the desired reaction product at its optimal rigidity.

A method to solve this problem may be the use of finger beam stops
in the dispersive plane. This was done in the HRIBF Recoil Mass
Separator \cite{CO92} to prevent high intensity beam components
from arriving at the final focal plane.

A method that improves both the separation and increases the
transmission is the {\it Gas-filled Mode} \cite{PA89}. Ions
passing through such a system will undergo frequent
charge-changing atomic collisions with the net effect of an
average charge q$_{aver}$ and therefore an average magnetic
rigidity. This can improve the transmission of ions through a
magnet system at the expense of a certain but often acceptable
increase of the emittance of the ion beam due to multiple
scattering. This method is particularly useful in fusion reactions
where the evaporation residues have a well defined velocity.
Typically one uses normal kinematics in these reactions to
guarantee complete separation when the fusion cross sections are
extremely small. For secondary beam production, however, when
clean separation is not essential and an ion catcher with large
acceptance can be employed, also inverse kinematics can be used.
We will come back to this discussion in section \ref{gas section}.
In the dual magnetic separator this mode can be realized by
operating {\it Section 2} in {\it Gas-filled Mode}. The target is
then located at T2.

\section{Ion optics}

Extensive first order ion-optical studies were conducted to design
a separator magnet system that could accommodate both {\it Fragment}
and {\it Gas-filled Mode}s. This led to the specifications summarized
in Table \ref{design parameters}. While the system is mainly designed
to accommodate the desired reaction products, also the ion optics of
the separated beam had to be considered for a variety of circumstances
to allow a clean separation and elimination of a high-intensity
primary beam. Constraints of fixed building and shielding structures
require that the separator has to be very compact without giving up
flexibility for the different operational modes and production
reactions.

Special effort was made to minimize the number of quadrupoles. An
initially considered quadrupole (Q3) between dipoles B1 and B2 could
be omitted. This was possible by adjusting the edge angles of B1 and
B2. It is, however, not possible to eliminate any of the present eight
quadrupoles without compromising one or the other design requirement
of the facility. As an example, in the {\it Fragmentation Mode}
quadrupole Q4 might be omitted at the cost of flexibility; however,
it is absolutely necessary in the {\it Gas-filled Mode} where
{\it Section 1} is used as beam line. This also requires reversing
the polarity of quadrupole Q4 and Q5 to obtain a small beam spot
at T2.

First-order design calculations of the beam and initial calculations
of the separator were performed using the TRANSPORT code \cite{RO78}.
The quadrupole gradients provided by the dipole edge angles, seen
in Fig. \ref{fig:Layout} were adjusted to reduce not only the number
of quadrupoles as mentioned above, but also to minimize the quadrupole
magnet strengths. This was crucial for minimizing the size and power
consumption of the large quadrupoles.

Due to the large angle and momentum acceptances, higher order
calculations were indispensable and were conducted using the COSY
Infinity code \cite{BERZ}. Aberrations up to 3$^{rd}$ order were
found that would have significantly affected the design parameters.
They were reduced by shaping the edge curves of the eight available
dipole ends as can be seen in Fig. \ref{fig:Layout}. In addition
a hexapole (HEX) was included between dipoles B3 and B4. This
hexapole improves the optics by reducing 2$^{nd}$ order aberrations
in the {\it Fragmentation Mode}.

\begin{figure}
\centering
\includegraphics[width=140mm,angle=0]{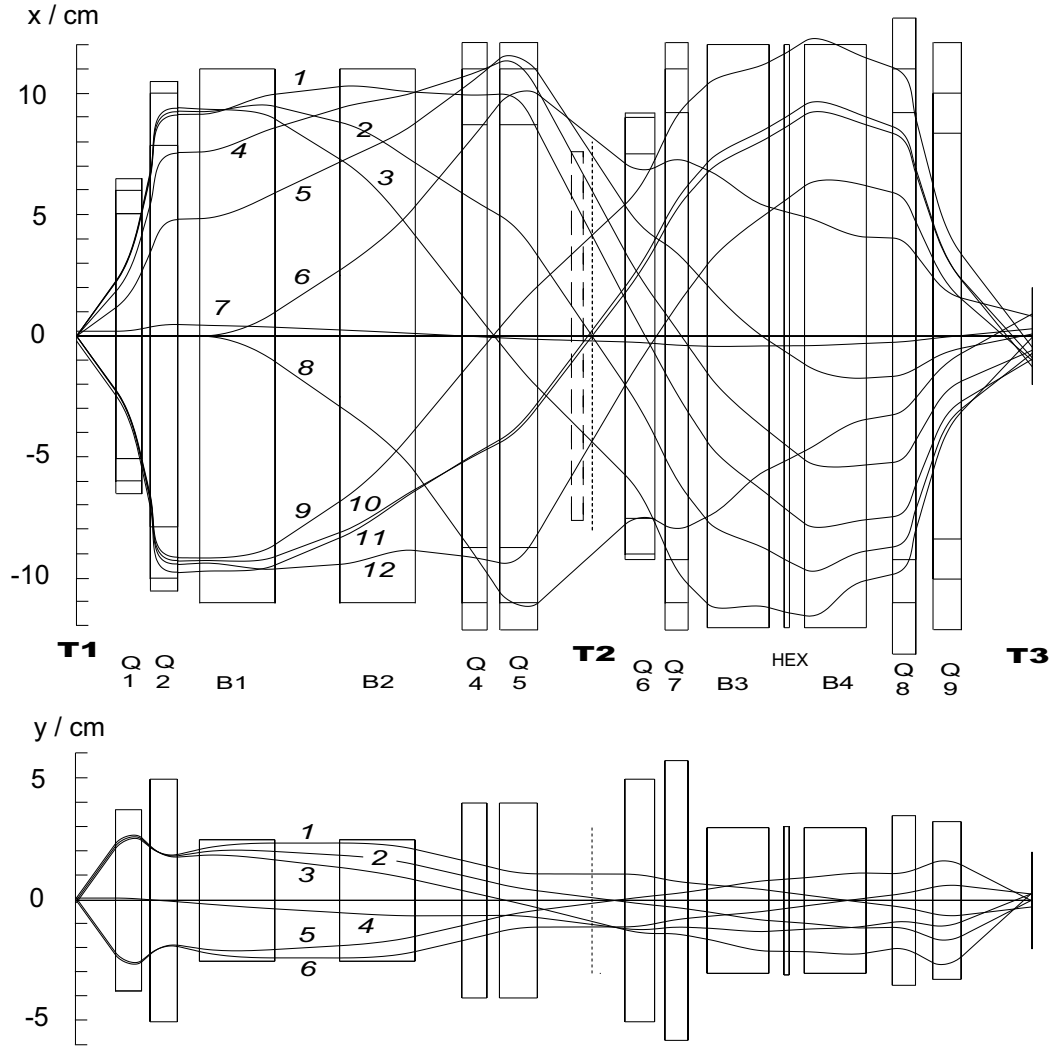}
\caption{ Ion optics of the TRI$\mu$P separator in {\it
Fragmentation Mode}. The starting values of the rays are listed in
Table \ref{raylist}. The notations refer to the
same elements of the layout shown in Fig. \ref{fig:Layout}. For
more details see text.} \label{fig:Optics_fragm}
\end{figure}

Results of the final ion-optical design calculations in 3$^{rd}$
order of the {\it Fragmentation Mode} in horizontal (x) and
vertical directions (y) along the central ray (z) are shown in
Fig. \ref{fig:Optics_fragm}. The system is 9.82 m long from the
target location T1 to the dispersive plane T2 (indicated by the
dotted line) and 18.20 m to the achromatic plane T3. The effective
field lengths (EFL) and good field regions of the four dipole
magnets B1 - B4 are indicated by the rectangles. For the
quadrupoles Q1 - Q9 the rectangles indicate in z direction the
EFL, while the transverse boundaries given by the three lines
indicate the dimensions, starting from the inside, of the pole
radius, the good field region, and the physical limits of the
vacuum chambers, respectively. The dashed box in front of T2
indicates a temporary horizontal limitation, which will be removed
later. To elucidate the optics, twelve horizontal and six vertical
rays characteristic for the separator operation are shown in Fig.
\ref{fig:Optics_fragm}, the starting values of the rays are listed
in Table \ref{raylist}.

\renewcommand{\arraystretch}{1.0}
\begin{table}
\caption{Starting values of the rays shown in Fig. \ref{fig:Optics_fragm}} \label{raylist}
\centering
\begin{tabular}{|c|c|c|c|c|c|}
  \hline
  ray & x [mm] & $\Theta$ [mrad]& $\Delta$E/E [\%]& y [mm] & $\Phi$ [mrad] \\
  \hline
  1 & 0 & 30 & 2.2 & -1 & 30 \\
  2 & 0 & 30 & 0 & 0 & 30 \\
  3 & 0 & 30 & -2.2 & 1 & 30 \\
  4 & 0 & 25 & 3.2 & 1 & 0 \\
  5 & 0 & 16 & 4.0 & 0 & -30 \\
  6 & 0 & 0 & 4.4 & 1 & -30 \\
  7 & 2 & 0 & 0 &  &  \\
  8 & 0 & 0 & -4.4 &  &  \\
  9 & 0 & -30 & 2.2 &  &  \\
  10 & 0 & -30 & 0 &  &  \\
  11 & -2 & -30 & 0 &  &  \\
  12 & 0 & -30 & -2.2 &  &  \\
  \hline
\end{tabular}
\end{table}

The three pairs of horizontal rays
({\it1}/{\it9}),({\it2}/{\it10}), and ({\it3}/{\it12}) show the
dispersive focal plane in T2 with an energy dispersion of 1.95
cm/$\%$. The rays {\it7} and {\it11} explore the effects of a
target size of $\pm$ 2 mm. The rays {\it1}, {\it4}, {\it5},
{\it6}, {\it8} and {\it12} ~ test the energy acceptance as
function of angle. At  0$^{\circ}$, particles with an energy
spread of $\pm$ 4.4 $\%$ will be accepted limited by the good
field region of quadrupole Q5. The accepted energy spread will
gradually decline with increasing angle. At the maximum accepted
angle of $\pm$ 30 mrad the maximum accepted energy spread is
$\pm$ 2.2\%.

All horizontal rays arrive at the focal plane T3 within a space of
20 mm, sufficiently small for the following degrader/ion catcher.
This was achieved by 2$^{nd}$ order corrections built into the
dipole entrances and exits and the hexapole HEX. This hexapole has
an effective field length of 150 mm and an aperture radius of 90
mm with a maximum pole tip field of 0.07 T. The horizontal image
size is minimized by an appropriate hexapole excitation. It will
increases to about 30 mm if the hexapole magnet is switched off.
The vertical ion optics is designed to keep all rays within the
gaps of 50 mm of the dipoles B1 and B2 and 60 mm of B3 and B4. The
focal plane angle at T2 is close to 90$^\circ$. The vertical
envelop at T2 is relatively large due to the vertical
magnification of -10.0 but a small vertical image of about $\pm$ 5
mm is achieved in the final focal plane T3 mainly due to the
magnification of 3.4 at this location. The horizontal
magnifications for Section 1 and the complete system are -0.95 and
1.8, respectively. The above calculations show that for full
transmission, target spot sizes of less than $\pm$ 2 mm horizontally
and $\pm$ 1 mm vertically are required.

\begin{center}
\begin{figure}
\begin{center}
\includegraphics[width=114mm,angle=0]{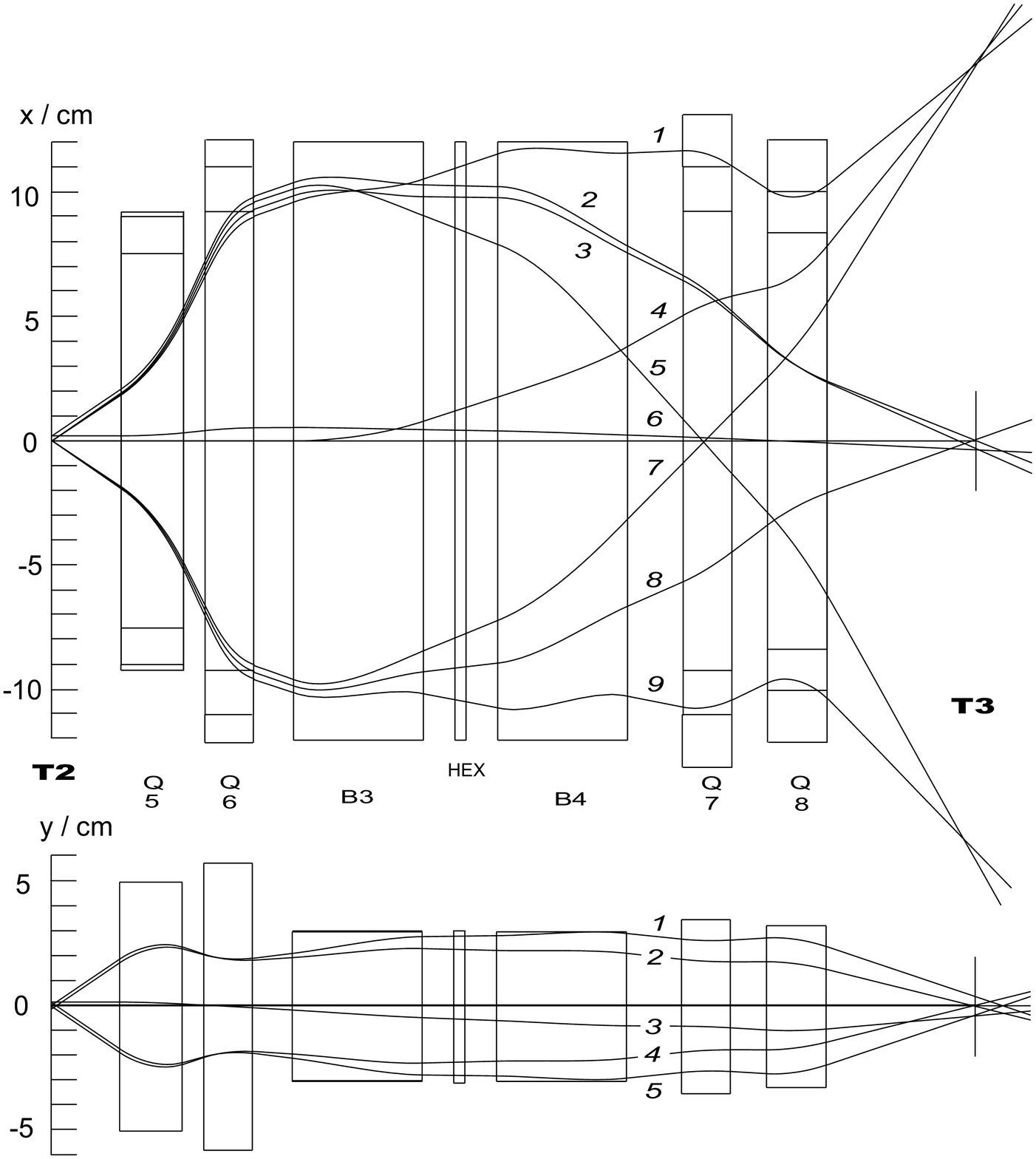}
\caption{ Ion optics of the TRI$\mu$P separator in {\it Gas-filled
Mode}. The starting values of the test rays are listed in Table
\ref{raylist2}. The notations refer to the same
elements of the layout shown in Fig. \ref{fig:Layout}. For more
details see text.} \label{fig:Optics_gas}
\end{center}
\end{figure}

\end{center}
\begin{table}
\caption{Starting values of the rays shown in Fig. \ref{fig:Optics_gas}} \label{raylist2}\centering
\begin{tabular}{|c|c|c|c|c|c|}
  \hline
  ray & x [mm] & $\Theta$ [mrad]& $\Delta$E/E [\%]& y [mm] & $\Phi$ [mrad] \\
  \hline
  1 & 0 & 30 & 4.0 & -1.5 & 30 \\
  2 & 2 & 30 & 0 & 0 & 30 \\
  3 & 0 & 30 & 0 & 1.5 & 0 \\
  4 & 0 & 0& 4.0 & 0 & -30 \\
  5 & 0 & 30 & -4.0 & 1.5 & -30 \\
  6 & 2 & 0 & 0 &  &  \\
  7 & 0 & -30 & 4.0 &  &  \\
  8 & 0 & -30 & 0 &  &  \\
  9 & 0 & -30 & -4.0 &  &  \\
  \hline
\end{tabular}
\end{table}

In the {\it Gas-filled Mode}, {\it Section 1}  serves as beam line
and the target (e.g. a carbon foil) is installed at T2. A thin
window (e.g. 2.5 $\mu$m HAVAR) is installed in front of the target
to separate the evacuated beam line and the gas-filled section.
The ion-optical design is shown in Fig. \ref{fig:Optics_gas}. This
is the initial magnet setting without gas filling. The ion optics
will be changed significantly when the system is gradually filled
with gas for optimum separation of beam and products since the
particles gradually decrease in energy along their path from T2 to
T3.

The upper panel shows the magnetic elements and nine selected rays
in the horizontal midplane. The lower panel shows five vertical
rays. The starting values of the test rays are listed in Table
\ref{raylist2}.

The three pairs of horizontal rays
({\it1}/{\it7}),({\it3}/{\it8}), and ({\it5}/{\it9}) show the
dispersive focal plane in T3 with an energy dispersion of 4.0
cm/$\%$. The rays {\it2} and {\it6} explore the effects of a
target size of $\pm$ 2 mm. Ray {\it4} shows a trace at 0$^{\circ}$
with an energy difference of  $+4.0 $ \%. All horizontal rays of a
certain momentum arrive at the focal plane T3 within 10 mm. This
was achieved by 2$^{nd}$ order corrections built into the dipole
entrances and exits as mentioned above. The vertical ion optics is
designed to keep all rays within the gaps of 60 mm of the dipoles
B3 and B4. The above calculations show that for full transmission,
target-spot sizes of $\pm$ 2 mm horizontally and $\pm$ 1.5 mm
vertically or smaller are required. The focal plane angle at T3 is
70$^\circ$ due to the concave curvatures of the effective field
boundaries of B4. This correction is not optimal for the {\it
Fragmentation Mode}, but the hexapole between B3 and B4 can be
used to minimize the beam-spot size at T3. As we will see below
the image size in this mode is dominated by statistical
charge-changing processes in the gas.

\section{Magnet Design}

The magnets are designed to provide the good field region required
in both transverse directions and the necessary field strength
according to the ion-optical calculations. All magnets are
operated by highly stabilized direct currents and are designed to
operate at fields below 1.65 Tm where iron saturation could be
kept small. All iron pole pieces and return yokes are therefore
machined of solid, soft iron. All coils are normal conducting with
hollow copper conductors to allow water cooling. The coil
temperatures are kept below 55$^\circ$ C. All magnets including
their vacuum chambers and supports were manufactured by commercial
vendors with significant experience in the design and construction
of similar magnet systems.

\subsection{Dipole Magnets}

\begin{center}
\begin{figure}
\begin{center}
\includegraphics[width=80mm,angle=0]{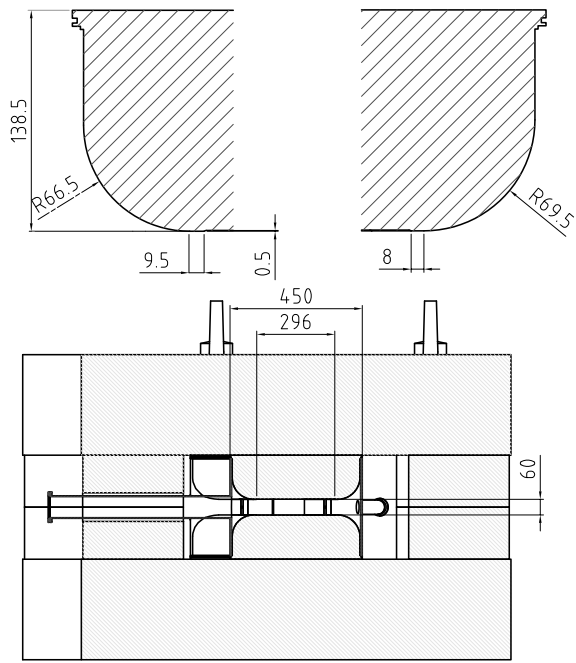}
\caption{ Cross section of the assembled dipole magnet B3 (bottom) and the pole shoe (top) with circular edges
and modified Rose shims indicated by the numerical values. All dimensions are in millimeter.} \label{fig:shoes}
\end{center}
\end{figure}
\end{center}

The four dipole magnets were manufactured by the Danfysik company
according to the specifications summarized in Table \ref{Tab2}.
All dipole magnets are H-type magnets. In order to achieve the
good field region for a minimum of magnet iron, two special
features are incorporated in the design:
\begin{itemize}
\item[(i)]The radial profile has circular side profiles with
modified Rose shims as shown by the measures in the upper part of
Fig. \ref{fig:shoes}. This allows for dipole B1 and B2 with a gap
of 50 mm, a horizontal good field region of 220 mm with a pole
width of only 400 mm. Dipole B3 and B4 with gaps of 60 mm have the
same characteristics with a slightly larger good field region of
240 mm, as required, and a pole width of 450 mm. This pole face
shaping saves some 30\% of the iron compared to magnets with
traditional Rose shims and homogeneous regions of the same size.

\item[(ii)] The amount of magnet iron is further reduced by a wrap-around chamber which is precision machined
and subsequently welded to the sides of the pole piece with special, thin welding lips to minimize heat effects
in the magnet iron. The accuracy of the gap is maintained by precision, non-magnetic spacers that are held with
bolts through the pole pieces. All spacers are protected from excessive beam and heat exposure by insulated
tungsten shields that allow to monitor any beam current hitting the shields.
\end{itemize}

\renewcommand{\arraystretch}{1.5}
\begin{table}
\begin{center}

\caption{Design parameters of the dipole magnets} \vspace{\baselineskip}

\begin{tabular}{|l|c|l|l|l|l|}
 \hline
Parameter & &\multicolumn{4} {|c|} {Dipole Magnet Type} \\
\hline
&& B1&B2&B3&B4\\
\hline
Bending radius  & mm & 2200 & 2200 &1800&1800\\
 \hline
Max. rigidity & Tm & 3.6 & 3.6 & 3.0 &3.0\\
 \hline
Max. magnetic field B & T & 1.64 & 1.64 & 1.67&1.67\\
\hline
Bending angle & deg &37.5 &37.5 &37.5 &37.5 \\
\hline
Central ray arc length &mm &1439.9 &1439.9 &1178.1 &1178.1 \\
\hline
Vertical gap, full size &mm &50 &50 &60 &60 \\
\hline
Good field region, dB/B $< \pm$ 0.02 $\%$ &mm &220 &220 &240 &240\\
\hline
Pole width &mm &400 &400 &450 &450\\
\hline
Entrance edge angle, vert. focusing &deg &18.75 &10.0 &0 &0\\
\hline
Entrance edge curvature, 1/radius &1/m&0.67&-0.2&0.88&-0.2\\
\hline
Exit edge angle, vert. focusing &deg &10.0 &18.75 &18.75 &18.75\\
\hline
Exit edge curvature, 1/radius & 1/m &0.0 &-1.29 &0.0 &-1.36\\
\hline
Max. current for magnet&A&380&380&380&380\\
\hline
Max. Voltage for magnet&V&90&90&80&80\\
\hline
Weight of iron&kg&11000&11000&11000&11000\\
\hline
Weight of coil&kg&600&600&500&500\\
 \hline
 \end{tabular}
\label{Tab2}
\end{center}
\end{table}

As shown in Fig. \ref{fig:Layout}, all magnets have 0$^\circ$ ports in both directions for viewing, alignment
and other access needs that may arise. Additional ports are provided in the middle of the magnets through the
inside return yoke. These ports were used for radial field maps with Hall probes and to obtain the magnetic flux
B versus current, B(I), with NMR probes immediately after assembly. Hall probes will be installed permanently to
allow setting of the fields without having to consider hysteresis and saturation that affect settings according
to coil currents.

The dipole design specifications in Table \ref{Tab2} include maximum current and voltage requirements to allow
the use of power supplies for the dipole magnets from other KVI facilities  that cannot run when the beam is
used in the TRI$\mu$P facility. For this reason dipole magnets B1 and B2 are powered in series by one power
supply.  Small correction coils are mounted on top of the main coils allowing for tuning differences between
both dipoles up to 3$\%$. The dipoles B3 and B4 are powered independently by using existing power supplies.

While 3-dimensional field calculations were performed by the
manufacturer to predict the end packs for the proper edge angles
and hexapole components, it was decided to machine the magnet ends
so that additional shims would allow small corrections. For this
purpose, the dipole magnets were assembled at first without
welding the vacuum chambers onto the pole pieces for access
reasons. The radial and B(I) field maps through the middle port
were performed as mentioned above. Subsequently field maps in the
midplane with a grid distance of 10 mm were performed in the
fringe field regions and in the middle of the magnets within a
section of $\pm$7$^{\circ}$ of the bend angle. In the measured
homogeneous part we verified that the field did not deviate more
than $2\times 10^{-4}$ from the average field. The effective field
boundary shapes were verified up to 2$^{nd}$ order and for the
entrance of B1 and B3 up to 3$^{rd}$ order. We required a
deviation less than 0.3 mm from the design shape without field
clamps. By adding shims on the end packs on either side of the
central ray, corrections between 0 mm and 2 mm brought the magnet
fields within specification with the exception of B4 where the
deviations exceeded slightly the 0.3 mm requirement. Higher order
COSY calculations were used to verify that the remaining
deviations did not jeopardize the resolving power design
specifications given in Table \ref{design parameters}. Some of the
eight end packs did not need corrections, a few needed one
iteration, and only in two cases a second iteration was necessary.
Final field maps for all dipole magnets were performed, documented
and are available for future purposes.

After a dipole magnet was optimized with this procedure, it was disassembled and the vacuum chamber welded in
place. In order to verify that this welding and reassembly procedure did not affect the original field map, we
mapped one magnet again, obviously with a somewhat reduced mapping region with the vacuum chamber in place. The
result showed an identical field map within the accuracy of the measuring method and well within specifications.
A small but measurable deviation was found in the fringe field region where the rather thick exit port is welded
in place. This deviation was within specifications.

\subsection{Quadrupole Magnets}
\begin{table}
\caption{Design parameters of quadrupoles.} \vspace{\baselineskip}
\begin{center}
\begin{tabular}{|l|c|l|l|l|l|l|l|l|l|}
 \hline
Parameter & &\multicolumn{8} {|c|} {Quadrupole Magnet Type} \\
\hline
&& Q1&Q2&Q4&Q5&Q6&Q7&Q8&Q9\\
\hline
Overall length & mm & 580 & 680 &630&880&680&680&700&700\\
 \hline
Focusing strength & T & 8.2 & 6.2 & 5.0 &6.3&5.6&2.8&2.0&3.3\\
 \hline
Eff. field length & mm & 480 & 550 & 500&780&550&420&420&500\\
\hline
Gradient &T/m &17.0 &11.3 &10.0 &8.1 &10.2 &6.7 &4.8 &6.7 \\
\hline
Horiz. good field  &mm &120 &200 &220 &220 &180 &220 &220 &200 \\
\hline
Aperture diameter &mm &100 &167 &184 &184 &150 &184 &184 &167 \\
\hline
Max.pole tip strength &T &0.85 &0.95 &0.88 &0.75 &0.77 &0.62 &0.45 &0.56\\
\hline
Current, magnet &A &256 &380 &398 &460 &360 &255 &160&255\\
\hline
Voltage, magnet &V &37 &107 &74 &62 &102 &78 &36&78\\
\hline
Power supply rating &A&270&400&420&480&380&270&180&280\\
&V&45 &115 &80 &70 &110 &90 &45&90\\
\hline
Magnet weight &kg &800 &3180 &1550 &4900 &3180 &1650 &630&1650\\
\hline
Inhomogeneity & &0.2\% &0.2\% &0.2\% &0.2\% &0.2\% &0.2\% &0.2\%&0.2\%\\
 \hline
 \end{tabular}
\label{Tab3}
\end{center}
\end{table}
\begin{figure}
\centering
\includegraphics[width=120mm,angle=0]{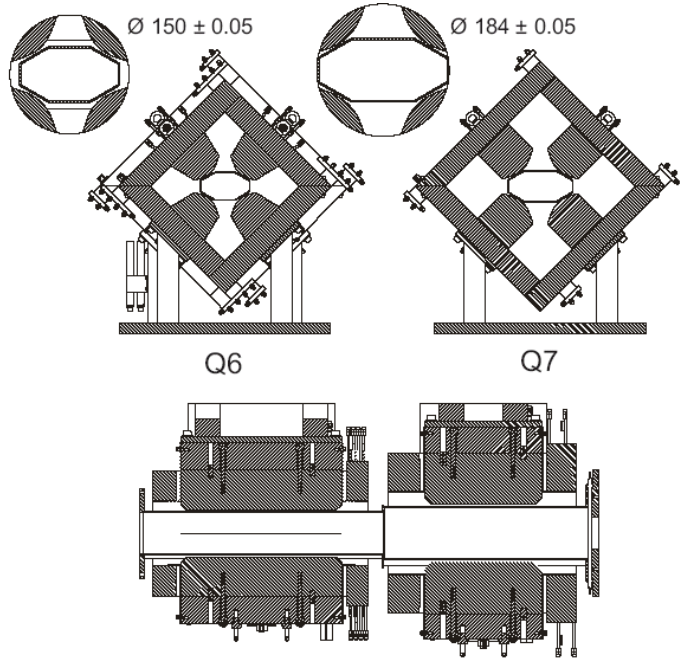}
\caption{ Assembly drawing of the quadrupole magnet doublet Q6, Q7
including their vacuum chamber. } \label{fig:Q}
\end{figure}

There are eight quadrupoles in the separator system, manufactured
by the company SigmaPhi according to the specifications in Table
\ref{Tab3}. The horizontal good field regions are required to be
significantly wider than the vertical ones, reflected in the shape
of the vacuum chambers. As an example this is shown in Fig.
\ref{fig:Q} for the quadrupole doublet Q6, Q7. All quadrupole
magnets were mapped and tested at the factory and delivered
including the vacuum chambers and supports. As in the case of the
dipole magnets, most of the electrical specifications were
determined to allow the use of existing power supplies except for
two power supplies that had to be purchased.

\section{Diagnostics and Instrumentation}

For the proper setup, optimization and operation of the separator a series of diagnostic elements is needed. A
variety of devices are required to tune and stop the beam, in addition standard Si solid state detectors were
used in the intermediate and focal plane to observe the fragments. Some of the instrumentation is discussed
below.

\subsection{Beam diagnostics and stops}

The target chambers T1 or T2 and the vacuum chamber between B1 and B2 are equipped with retractable harps. The
harps consist of 24 wires that are stretched in horizontal and vertical directions with 1 mm or 2 mm spacing.
Signals from the wires are individually read out and visualized as a beam profile in vertical and horizontal
directions. The harps are standard equipment for beam tuning in the beam lines \cite{SC95}. They can be operated
to tune with the full beam and viewed simultaneously in the setup phase of an experiment. In T1 and T2 target
ladders are used with scintillating targets, they can be viewed by means of CCD cameras through plexiglass
ports. The lifetime of these viewing targets (Zn$_2$SiO$_4$ on Al foil) is limited. They are  mostly used for the final
tuning. Both snapshots of the harp data and viewing targets can be stored for reference. In the commissioning
phase scintillating screens were also used on a movable platform in a focal-plane chamber T3.

In anticipation of high beam currents, the beam stops in front of T1 and T2 have been designed to allow water
cooling. When tuning for fragments the beam is stopped on one of the slits located at  SH2, SH3, or SH4 for the {\it
Fragmentation Mode} and SH5 or SH6 for the {\it Gas-filled Mode}.

At present there are three pairs of horizontally movable slits
that are remotely controlled and can be installed at any of the
mentioned locations depending on the requirements of the
experiment. The movable slits are also used to verify the ion
optics, e.g. focus conditions in conjunction with the beam or
other setup conditions.

\subsection{Particle identification}

In order to optimize the separator for transport and focusing of the desired reaction products at the exit of
the separator, it is important to have online particle identification. While the optimal detection system
depends on particle type, we have used a 100 $\mu$m thick Si detector with a diameter of 20 mm that provided
either energy E or energy loss $\Delta$E and a timing signal. As we will see below, this detection is well
suited for identifying light to medium-mass particles. One detector is permanently mounted on a movable arm
(horizontal plane) in T2. During the initial tests and development runs, a second Si detector was mounted on a
movable table at the focal plane T3.

\subsection{Gas target}

The {\it Fragmentation Mode} has been tested with the $^1$H($^{21}$Ne, $^{21}$Na)n reaction.   For initial tests
with low beam intensities, thin polyethylene foils as hydrogen targets were used, with carbon as the only other
target component. The hydrogen content of these targets, however, will diminish rapidly with higher ionization
density, i.e. when the beam current is increased or for low energy and/or heavy element beams.

For higher beam currents, a hydrogen-filled gas target is used. This target has 2.5 $\mu$m HAVAR foils  as
windows following a design of a target in use at Texas A\&M \cite{GO04}. In order to increase the density by
about a factor of 4, the gas cell is cooled down to liquid nitrogen temperature. The gas cell is approximately
10 cm long with windows of 1.25 cm diameter. It can be moved vertically to allow insertion of a target ladder at
the object position in T1. The gas target and its accessories were designed and produced at North Carolina State
University \cite{YO05}.

\subsection{Separation by differential stopping}

In {\it Fragmentation Mode}, the separator {\it Section 1}
provides a dispersive focal plane, where the beam - if not stopped
previously - and other undesirable products can be separated by
momentum selection using horizontal slits. The design momentum
acceptance of $\pm$ 2$\%$ translates in a 16 cm wide focal plane
due to the momentum dispersion of about 4 cm/$\%$. Magnet {\it
Section 2} reverses the effect of the dispersive function of {\it
Section 1} and provides an achromatic beam spot of the order of 20
mm diameter, well suited to be accepted in a following ion
catcher.

However, all particles, including the undesirable ones, that pass
through the momentum slits will also enter the ion catcher. For
further separation a degrader can be used.  Such a degrader will
provide mass selection, because of the differences in energy loss.
Because of the momentum acceptance of $\pm$ 2$\%$ also the desired
products will be dispersed, effectively disturbing the
achromaticity. This effect can be eliminated by changing the
thickness of the degrader along the dispersive plane.
Such a degrader is called a "wedge" and can be
realized by bending a foil of constant thickness. A holder for
bent foils with dimensions corresponding to the intermediate plane
was made.

\section{Installation and commissioning}

In November 2003 the modifications of shielding and the installation of power and utility lines in the
designated TRI$\mu$P separator hall (T-cell) was completed. At this time all major components, in particular the
magnets and their support, had been delivered. Only two additional power supplies had to be purchased as most of
the magnets were designed to be able to use existing power supplies not operated at the same time as the
TRI$\mu$P separator.

Installation of the separator itself started in December 2003 and was completed in April 2004 followed by
commissioning runs in May 2004. We tested the basic ion-optical parameters using a 43 MeV/nucleon
$^{21}$Ne$^{7+}$ beam (B$\rho$ = 2.86 Tm). A nearly circular primary beam spot of about 2 mm diameter (FWHM) was
achieved at T1. This beam was transmitted into the dispersive focal plane at T2 where it had dimensions of 5 mm
by 8 mm in the horizontal and vertical directions. The beam was subsequently put onto a viewer at T3, where the
achromatic image size was approximately 10 mm by 10 mm in both transverse directions. In order to verify the
momentum dispersion and to obtain information about the momentum acceptance, copper foils of 7.5, 15, and 22.5
$\mu$m were inserted at the target location T1. The displacement at T2 was measured using a large Zn$_2$SiO$_4$ viewer
marked with a centimeter scale. The momentum loss of the beam in a 22.5 $\mu$m foil was 1.0$\%$. The beam
consists of fully stripped $^{21}$Ne$^{10+}$ ions after the target. The momentum dispersion measured in the
dispersive focal plane (T2) was 4.2 cm/$\%$ corresponding to 2.1 cm/$\%$ energy dispersion in non-relativistic
approximation. This is in agreement with the design calculations. A circular beam spot of about 20 mm diameter
(FWHM) was observed in the achromatic final focal plane (T3).

The ion optics of the {\it Gas-filled Mode} was tested by
transporting the $^{21}$Ne$^{7+}$ beam onto the target T2. Without
gas filling the momentum dispersion in the final focal plane was
measured by inserting the copper foils and using the method
mentioned above. The resulting momentum dispersion of 6.9 cm/$\%$
is in agreement with the design calculations. After having
verified the basic optical design parameter and that all separator
components worked satisfactorily, we proceeded to produce
$^{21}$Na isotopes required for one of the initial experiments.

\section{Production of $^{21}$Na in \textit{Fragmentation Mode}}

After successful commissioning and verification of ion optics and
design parameters, we produced $^{21}$Na using the (p,n) reaction
in inverse kinematics with a $^{21}$Ne$^{7+}$ beam at 43
MeV/nucleon. A 20 mg/cm$^2$ polyethylene (CH$_2$)$_n$ foil was
used in setup runs where low beam intensity of a few nA current
was sufficient. For higher beam intensities of up to 30 particle
nA we used the LN$_2$ cooled hydrogen target at 1 atm. The
emerging $^{21}$Ne$^{10+}$beam with a magnetic rigidity B$\rho$
approximately 9$\%$ higher than the desired $^{21}$Na$^{11+}$
isotopes is stopped in the movable beam stop SH2 between the first
dipole magnets B1 and B2. The movable arm with the 100 $\mu$m
thick, circular silicon detector of 20 mm diameter is installed
immediately downstream of the wedge in chamber T2. A second
silicon detector of 150 $\mu$m thickness was installed in the
focal plane T3 also on a platform that moves along the horizontal
axis.

\begin{figure}
\centering
\includegraphics[width=140mm,angle=0]{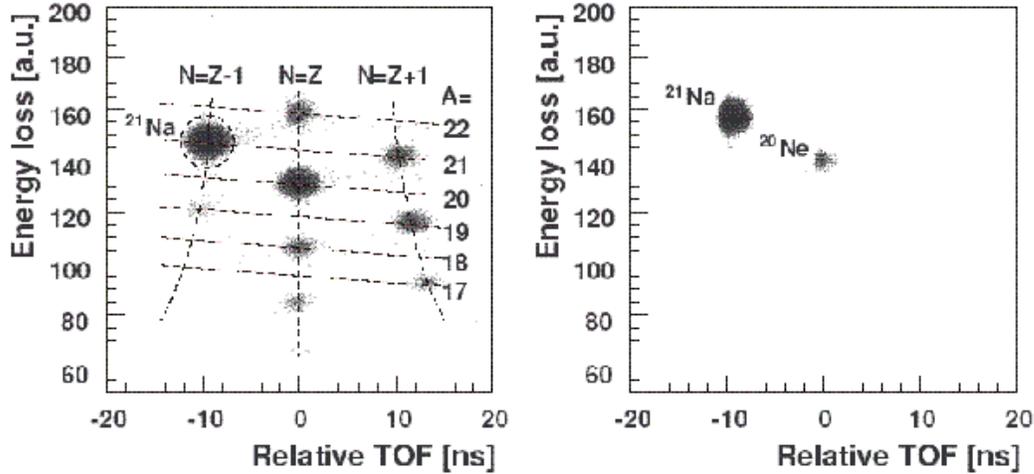}
\caption{ Left: The production of $^{21}$Na. The time of flight vs. energy loss in a 100 $\mu$m silicon detector
shows various nuclides at the intermediate plane (T2). Right: At the final focal plane (T3) there is only
$^{21}$Na and a small contamination of stable $^{20}$Ne  which could be reduced to below 0.5 \%. }
\label{fig:21Na}
\end{figure}

The left side panel of Fig. \ref{fig:21Na} shows the energy loss
of particles in the detector at T2 vs. the time-of-flight (TOF)
relative to the Radio-Frequency (RF) phase of the cyclotron. The
RF was 32 MHz, providing a dynamic range of 31 ns for the TOF
measurement,  just sufficient to observe the dynamic range of
particles in the Ne mass region without ``wrap-around'' that would
complicate particle identification. Due to the nature of the
reaction the $^{21}$Na products are strongly focused and most of
these fragments pass through the detector, i.e. the setting of the
magnetic fields in {\it Section 1} is adjusted to optimize the
$^{21}$Na yield. The detector is left in place during the
measurements at T3, functioning as a uniform degrader. In the
panel on the right hand side the same spectrum is shown but this
time with a gate on particles arriving in the detector at the
focal plane T3 that has also a 20 mm diameter. Optimizing the
rigidity of {\it Section 2}, one can tune for the maximum yield of
$^{21}$Na. Only a small contribution of 0.5\% of $^{20}$Ne is
remaining while all other impurities including beam particles are
reduced to insignificant contributions. Note that in this test run
the left side spectrum was measured using a polyethylene foil,
while the right side was measured with the cooled hydrogen target
to allow higher beam and therefore higher production rates.
$3.2\times 10^3$ $^{21}$Na/particle-nA /s of primary beam was
already achieved. In these and other production measurements we
profited greatly from using the code LISE++ of Tarasov and Bazin
\cite{BA02,TA02} which allows to enter the specification of the
separator. Unfortunately, it does not yet work for direct
reactions, so that the fragmentation  or fusion-evaporation mode
of the program has to be used for initial settings of the
separator. Recently we produced $^{22}$Mg, $^{20}$Na, $^{19}$Ne
and $^{12}$N. All using inverse (p,n) reactions except for
$^{22}$Mg where a $^{23}$Na beam was used in a (p,2n) reaction at
32 MeV/nucleon. Production of $^{20}$Na and $^{19}$Ne was
optimized resulting in a yield of $10^3$ and $10^4$/s/particle-nA,
respectively.

For 1 kW beam the $^{21}$Ne mentioned above corresponds to 1.1
particle-$\mu$A.  The maximum useful target thickness is obtained
when the energy loss difference of the primary beam and the
secondary beam corresponds to the energy acceptance (8 \%) of the
separator. Note, that using a H$_2$ target the pressure windows
only affect the acceptance due to straggling but not due to energy
loss. In this way using a 10 bar target one may reach 10$^8$
$^{21}$Na particles/s.

This clean secondary $^{21}$Na beam was used later in a first experiment in collaboration with a group from LPC,
Caen, France \cite{Achouri} to study the $\beta$-decay branching ratio in the $^{21}$Na decay. This ratio is
important for the interpretation of the $\beta$-$\nu$ correlation measurements \cite{SC04} performed as tests of
the Standard Model.

\section{Tests of \textit{Gas-filled Mode}}\label{gas section}

The planned physics program \cite{JU05,BE03} requires also the production of very heavy isotopes like $^{213}$Ra
that may be produced by using the fusion-evaporation reaction in inverse kinematics. If the radium atom  has
nearly degenerate states of opposite parity, $^{213}$Ra could be a candidate for the search for a (forbidden)
permanent electric dipole moment.

\begin{figure}
\centering
\includegraphics[width=0.6\textwidth,angle=0]{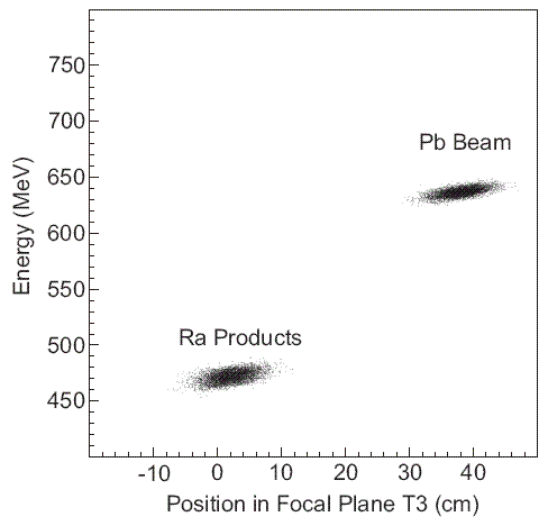}
\caption{Simulation of the separation of Ra produced by a 7 MeV/nucleon Pb beam impinging on a carbon target. In this mode separator {\it Section 2} was filled with 5 mbar Ar.} \label{fig:Pb-Ra}
\end{figure}

In an early phase of the design, calculations were made with the transport code from M. Paul \cite{PA89}, which
incorporates charge-changing atomic collisions and energy loss in a gas-filled magnetic system. In particular
the reaction $^{206}$Pb on $^{12}$C was considered where one could collect all Ra residues with 100\%
efficiency, as the effect of the evaporated neutrons is negligible compared to broadening due to multiple
scattering in the gas. The simulations were made to find whether the residues and the beam can be separated
sufficiently. In this application a clean separation is not essential if the ion-catcher device can handle the
additional load due to the beam contamination. In Fig. \ref{fig:Pb-Ra} the result of this calculation is shown
for a Pb beam of 7 MeV/nucleon on a carbon target placed at T2 with {\it Section 2} filled with  5 mbar Ar. Note
that the separation increases with increasing pressure due to the difference in stopping power. The separation
is about 13$\sigma$, where $\sigma$ is the Gaussian width of the position distribution of Pb; the separation
expressed this way is not sensitive to the pressure. The Gaussian width found in the simulation shown in Fig.
\ref{fig:Pb-Ra} is 2.5 cm, i.e., nearly 6 cm Full Width at Half Maximum (FWHM).

\begin{figure}
\centering
\includegraphics[width=0.6\textwidth,angle=0]{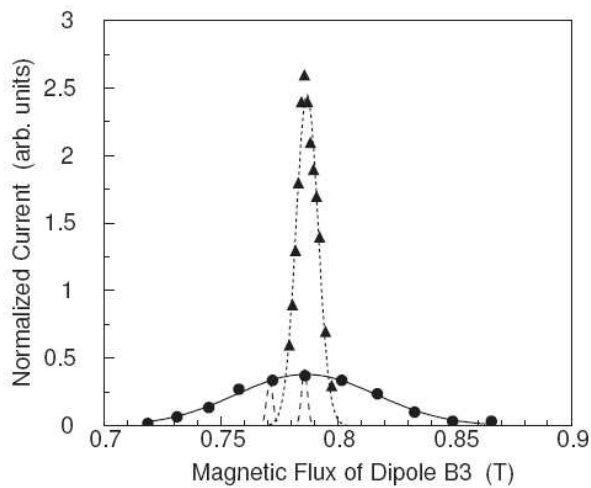}
\caption{ Experimental charge-state distribution of a $^{206}$Pb beam (full circles) are collected in a single
8.6 cm wide beam spot (full triangles); for further details see text. } \label{fig:Pb}
\end{figure}

To test some of the results a measurement was made with a $^{206}$Pb$^{29+}$ beam of 8.4 MeV/nucleon passing
through a HAVAR window of 2.5 $\mu$m thickness before impinging on a carbon target of 4.2 mg/cm$^2$ thickness.
Results are shown in Fig. \ref{fig:Pb}. Without gas filling the charge distribution after passing the foil and
the target is given by the full circles. A total of eleven charge states with a maximum intensity estimated to be at
q=60 were observed. The line through the points is a Gaussian distribution with a FWHM of 9\% $\delta$p/p. A
more detailed measurement over two charge states is indicated by the two dashed Gaussian curves that have each a
FWHM of 0.37\% $\delta$p/p. This is the overall resolution of the beam, including straggling in the entrance
foil in front of the target. The triangles represent a scan over the focal plane after filling the complete
system after the entrance foil with 2.5 mbar Argon gas. The eleven separated charge states are now concentrated in a
single peak corresponding to an average charge state due to the statistical process of charge-state changes in
the gas. For comparison of the data sets the measured setting of the data obtained with gas filling were shifted
in momentum to coincide with the maximum of the charge distribution measured in vacuum. For the same reason the
beam current data were scaled to obtain the same total integrated particle current. The resulting distribution
is well reproduced by a Gaussian distribution, as can be observed from the fitted dotted curve. The FWHM was
1.4\% $\delta$p/p or 8.6 cm in absolute width, i.e. nearly 4 times the width of a single charge state, but 6
times smaller than the full charge distribution. Also the dependence on Ar gas pressure was investigated. Since
the ion optics changes due to the energy losses in the gas, after each pressure change the magnet settings had
to be adjusted empirically for best resolution. The width depended little on pressure. The best resolution was
found at 2.0 mbar with a FWHM of 6.4 cm. At 5 mbar, the pressure corresponding to the calculations in Fig.
\ref{fig:Pb-Ra}, the FWHM was 7.7 cm, i.e. 30\% larger than the calculated value. The low beam current (0.2
particle-nA) did not allow us to conclude anything about the radium production or distribution. However, the
observations so far indicate that the calculated properties appear to apply. To collect all heavy reaction
products in {\it Gas-filled Mode} an ion catcher of about 10 cm diameter would be required.

\section{Summary}

A versatile magnetic separator for the effective collection and separation of radio isotopes produced with a
variety of heavy ion beams from the AGOR superconducting cyclotron was designed, built and commissioned at the
KVI as part of the TRI$\mu$P facility. The major magnetic and system components were manufactured by commercial
vendors. The system consists of two sections with a total of four dipole and eight large quadrupole magnets to produce
a dispersive intermediate and an achromatic final focus.  By shaping the dipole edges the system is corrected to
3$^{rd}$ order and the focal plane angles could be optimized. Aberrations up to 4$^{th}$ order were calculated.
It was found that orders higher than three were negligible. Special care was taken to include optimally the
existing infrastructure and hardware of the laboratory.

The maximum design rigidities of the separator
are given by the maximum beam rigidity of 3.6 Tm {\it (Section 1)}
and a maximum product rigidity of 3.0 Tm {\it(Section 2)}. The
angle  acceptances are 30 mrad in both transverse directions. The
momentum acceptance is $\pm$2.0$\%$. This allows to transmit a
large fraction of the phase space of the reaction products of
interest using the inverse reaction technique into a cooler
device. Extrapolating the yields for light products up to $10^8$
particles/s appears feasible.

The system is designed to separate reaction products from light (e.g. $^{21}$Ne) to very heavy $^{208}$Pb beams
with maximum beam intensities given by the power dissipation limit of about 1 kW in the cyclotron deflector.
{\it Section 1} of the system separates beam and reaction products  using B$\rho$ analysis alone. This works
particularly well for light ions and high energies where beam ions and products are nearly fully stripped. After
a degrader which differentiates different isotopes with similar B$\rho$, {\it Section 2} is used to provide a
relatively small achromatic image of the secondary beam. The second part of the separator also allows gas-filled
operation to collect heavy isotopes that have wide charge-state distributions emerging from the production
target located between the two sections.

During commissioning the system design parameters were established and a first clean isotope separation was made
in case of $^{21}$Na production. Initial tests of the {\it Gas-filled Mode} were made.

\section{Acknowledgments}
This work was supported by the Dutch Stichting vor Fundamenteel
Onderzoek der Materie (FOM) under program 48 (TRI$\mu$P) and the
EU RTD networks ION CATCHER and NIPNET, HPRI-2001-50045 and 50034
respectively. We are indebted to A. Young (TUNL) for providing a
liquid nitrogen cooled hydrogen gas target. The TRI$\mu$P group
would like to express their special thanks to H. Kiewiet, L.
Slatius and J. Sijbring  for their technical contributions. Our
gratitude also goes to the members of the AGOR cyclotron group and
the KVI support staff for their efforts in realizing this project.

\end{document}